\documentclass[12pt]{iopart}

\usepackage{iopams}
\usepackage{epsf}
\usepackage{cite}

\newcommand{\bd}{\begin{displaymath}}
\newcommand{\ed}{\end{displaymath}}
\newcommand{\be}{\begin{equation}}
\newcommand{\ee}{\end{equation}}
\newcommand{\ba}{\begin{eqnarray}}
\newcommand{\ea}{\end{eqnarray}}

\begin{document}

\paper[A trajectory-based understanding of quantum interference]
{A trajectory-based understanding of quantum interference}

\author{A S Sanz and S Miret-Art\'es}

\address{Instituto de F\'{\i}sica Fundamental\\
Consejo Superior de Investigaciones Cient\'{\i}ficas\\
Serrano 123, 28006 Madrid, Spain}

\eads{\mailto{asanz@imaff.cfmac.csic.es},
\mailto{s.miret@imaff.cfmac.csic.es}}

\begin{abstract}
Interference is one of the most fundamental features which
characterizes quantum systems.
Here we provide an exhaustive analysis of the interfere dynamics
associated with wave-packet superpositions from both the standard
quantum-mechanical perspective and the Bohmian one.
From this analysis, clear and insightful pictures of the physics
involved in this kind of processes are obtained, which are of general
validity (i.e., regardless of the type of wave packets considered) in
the understanding of more complex cases where interference is crucial
(e.g., scattering problems, slit diffraction, quantum control
scenarios or, even, multipartite interactions).
In particular, we show how problems involving wave-packet interference
can be mapped onto problems of wave packets scattered off potential
barriers.
\end{abstract}




\maketitle


\section{Introduction}
 \label{sec1}

Over the last 15 years or so, fields such as the quantum information
theory \cite{macchiavello,nielsen}, quantum computation
\cite{macchiavello,nielsen} and quantum control \cite{paul}
have undergone a fast development.
Simultaneously, due to the relevant role that {\it entanglement}
plays in all these fields, the idea of this phenomenon being the most
distinctive feature of quantum mechanics has also grown in importance,
although this already infers from earlier works \cite{schro}.
However, despite this relevance, in our opinion {\it quantum
interference} (together with diffraction) can still be considered
of more fundamental importance within the quantum theory, its most
striking manifestation being, most surely, the two-slit experiment,
which ``has in it the heart of quantum mechanics. In reality, it
contains the {\it only} mystery [\ldots]'', quoting Feynman
\cite{feyn}.
Quantum interference is the direct, observable consequence of the
{\it coherent} superposition of (quantum) probability fields.
This is precisely what is special about it: interferences in quantum
mechanics are not associated with or produced by a sudden transfer of
energy (in the way of a perturbation) along a material medium, as
happens with classical waves.
It is worth stressing that, apart from its importance at a conceptual
and fundamental level, quantum interference is also involved in a very
wide range of experimental situations and applications.
SQUIDs or superconducting quantum interference devices \cite{scalapino},
the coherent control of chemical reactions \cite{paul}, atom and
molecular interferometry \cite{berman} (in particular, with BECs
\cite{pritchard,chapman,alon}, where different techniques to recombine
the split beams are used \cite{haensel,hinds,andersson,kapale}) or
Talbot/Talbot-Lau interferometry with relatively heavy particles
(e.g., Na atoms \cite{chapman2} and BECs \cite{deng}) are just some
examples.
Moreover, it is also important to notice the role played by
interference when dealing with multipartite entangled systems as an
indicator of the loss of coherence induced by the interaction between
the different subsystems, for instance.

\begin{figure}
 \begin{center}
 \epsfxsize=9cm {\epsfbox{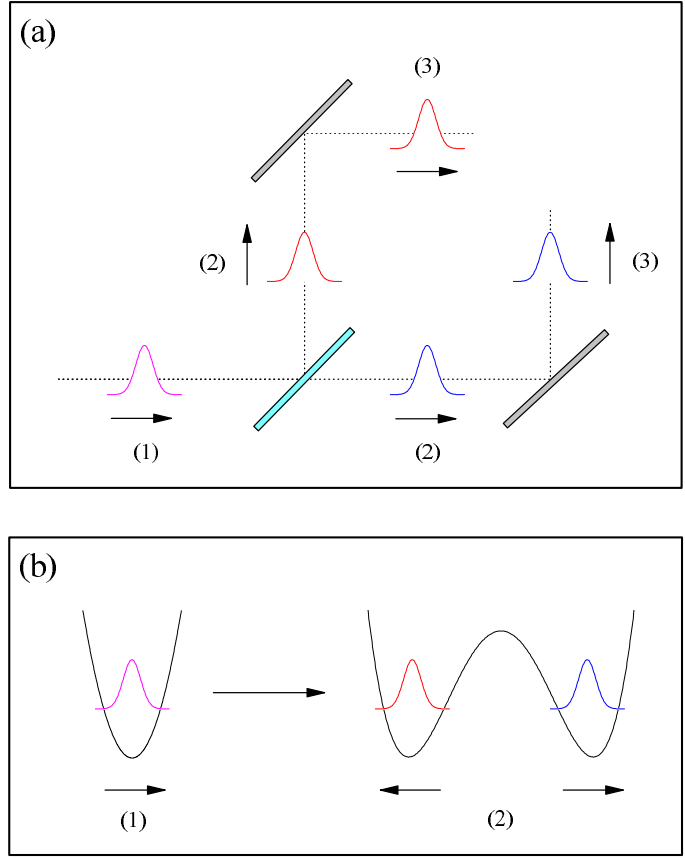}}
 \caption{\label{fig1}
  Two situations which give rise to a description of the wave function
  as a superposition of two wave packets at a certain time.
  (a) After reaching a beam-splitter, an incoming wave packet (1)
  splits into two ones (2).
  These wave packets move apart until they reach two reflectant
  mirrors, which redirect their propagation (3).
  The wave function describing the evolution of these two wave
  packets is a superposition which will display interference when they
  ``collide'' after some time in future (this fourth step has not been
  represented).
  (b) A wave packet confined in a potential well (1) is split into two
  wave packets by creating a barrier inside the potential (2) and then
  pulling outwards the two new potential wells (each containing a new
  wave packet).
  If the potential is turned off and the wave packets pulled inwards,
  the (superposition) wave function that describes them will again
  display interference (again, as in the previous case, we have not
  represented this step), as can be seen in BEC interferometric
  experiments.}
 \end{center}
\end{figure}

The actual understanding and interpretation of interference phenomena
arises from the, somehow, physical ``reality'' ascribed to the
superposition principle.
According to this (mathematical) principle, wave fields can be
decomposed and then recombined again to produce and explain the
interference patterns observed.
However, in Nature, wave fields constitute a whole and, therefore, it
is important to determine what this behavior means when we are dealing
with quantum phenomena displaying interference.
Here we analyze such consequences in the case of coherent
superpositions of Gaussian wave packets by means of properties which
can be derived from standard quantum mechanics (but that, to our
knowledge, are not considered at all in the literature) and also
from Bohmian mechanics \cite{bohm,holland-book}.
In the latter case, we will show that there exists a sort of mapping
which allows us to translate wave-packet interference problems onto
problems of single wave packets scattered off by potential barriers.
The interest in this kind of superpositions arises from the fact that,
rather than simple academical examples whose interference dynamics
is well known in the literature \cite{sanz-jpcm,sanz-talbot}, such
superpositions are experimentally realizable (and, indeed, used) in
atom interferometry \cite{haensel,hinds,andersson,kapale}.
In Fig.~\ref{fig1}, for instance, we have sketched two types of
experiments where coherent superpositions of Gaussian wave packets
can be produced.
Moreover, it is also important to highlight the interest that the
analysis presented here can have to better understand scattering
processes, quantum control scenarios or many-particle interactions,
where interference plays a fundamental role, as well as to design,
develop, improve and/or implement new trajectory-based algorithms
\cite{wyatt-book}, where one of the main drawbacks is the presence
of nodes in the wave function produced by interferences.

The organization of this paper is as follows.
To be self-contained, in section~\ref{sec2} we provide a brief
overview of quantum interference and Bohmian mechanics, as well as some
interesting properties which will be used and further discussed in next
section.
In section~\ref{sec3} we present the analysis of wave-packet
superposition dynamics and the main results derived from it.
Thus, we begin with a brief overview on (single) wave-packet dynamics
and its properties, and then we introduce our non-standard analysis
of two wave-packet interference processes and the analogy of this
problem with wave-packet scattering off barriers.
Finally, in section~\ref{sec4}, we provide a summary of the main
conclusions in this work as well as a discussion on their interest
in more complex problems.


\section{Interference and quantum trajectories}
 \label{sec2}

Since the Schr\"odinger equation is linear, it satisfies the
superposition principle.
Thus, given $\psi_1$ and $\psi_2$ satisfying separately this equation,
its superposition,
\be
 \Psi({\bf r},t) = \psi_1({\bf r},t) + \psi_2({\bf r},t) ,
 \label{quantum4}
\ee
will also be a valid solution.
Now, the wave amplitude $\Psi$ is not an {\it observable} magnitude,
but the probability density, $\rho = |\Psi|^2$, which provides a
statistical description of the system \cite{born}.
Due to the connection between $\Psi$ and $\rho$, it is clear that the
superposition principle does not hold for the latter.
Note that, after expressing $\psi_i$ ($i=1,2$) in polar form, i.e.,
$\psi_i = \rho_i^{1/2} e^{iS_i/\hbar}$, we find
\be
 \rho = \rho_1 + \rho_2 + 2 \sqrt{\rho_1 \rho_2} \cos \varphi ,
 \label{quantum5}
\ee
where $\varphi = (S_2 - S_1)/\hbar$.
The interference term in Eq.~(\ref{quantum5}) connects {\it coherently}
the probability densities ($\rho_1$ and $\rho_2$) related to each
partial wave ($\psi_1$ and $\psi_2$, respectively).

Since $\rho$ describes the statistical distribution associated with
a particle ensemble regardless of whether there is an interaction
potential connecting them or not, one can define an associate
probability current density as
\be
 {\bf J} = \frac{1}{m} \ \!
  {\rm Re} \left[ \Psi^* \hat{\bf p} \Psi \right]
  = - \frac{i\hbar}{2m}
  \left[\Psi^* \nabla \Psi - \Psi \nabla \Psi^*\right] ,
 \label{quantum7}
\ee
which indicates the flow of such an ensemble, with $\hat{\bf p} =
-i\hbar\nabla$ being the momentum (vector) operator.
Substituting (\ref{quantum4}) into (\ref{quantum7}), with the $\psi_i$
expressed in polar form, yields
\be
 \fl
 {\bf J} = \frac{1}{m}
  \Big[ \rho_1 \nabla S_1 + \rho_2 \nabla S_2
  + \sqrt{\rho_1 \rho_2} \nabla (S_1 + S_2) \cos \varphi
 + \hbar \ \! (\rho_1^{1/2} \nabla \rho_2^{1/2}
         - \rho_2^{1/2} \nabla \rho_1^{1/2}) \sin \varphi \Big] ,
 \label{quantum8}
\ee
which clearly shows that the superposition principle does not hold
either for ${\bf J}$.
The two magnitudes $\rho$ and ${\bf J}$ are related through the
continuity equation,
\be
 \frac{\partial \rho}{\partial t} + \nabla {\bf J} = 0 ,
 \label{quantum9}
\ee
which can be easily derived from the Schr\"odinger equation after
multiplying both sides by $\Psi^*$, adding to the resulting equation
its complex conjugate, and then rearranging terms.

Within this hydrodynamical picture of quantum mechanics \cite{madelung},
one can always further proceed as in classical mechanics and determine
the probability streamlines, i.e., the lines along which the
probability flows.
Consider the same polar ansatz used above, but for the total wave
function, $\Psi = \rho^{1/2} e^{iS/\hbar}$.
After substituting it into the Schr\"odinger equation and rearranging
terms, one obtains
\ba
 \frac{\partial S}{\partial t} & + &
  \frac{(\nabla S)^2}{2m} + V + Q = 0
 \label{eq14} \\
 \frac{\partial \rho}{\partial t} & + &
  \frac{1}{m} \ \! \nabla (\rho \nabla S) = 0
  \label{eq13}
\ea
from the real and imaginary parts of the resulting expression,
respectively.
Equation (\ref{eq13}) is the continuity equation (\ref{quantum9}),
with ${\bf J} = (\rho \nabla S)/m$.
Equation (\ref{eq14}), more interesting from a dynamical viewpoint, is
the quantum Hamilton-Jacobi equation, which allows us to reinterpret
the whole quantum-mechanical formalism in terms of the eventual paths
that the system can pursue when it is considered as a particle.
These paths are defined as solutions of the equation of motion
\be
 \dot{\bf r} = \frac{\nabla S}{m} = \frac{{\bf J}}{\rho}
  = -\frac{i\hbar}{2m}
   \frac{\Psi^* \nabla \Psi - \Psi \nabla \Psi^*}{\Psi^* \Psi} ,
 \label{eq16}
\ee
where $S$ represents a quantum generalized action.
Note that considering Eq.~(\ref{eq16}) means introducing a new
conceptual element into quantum mechanics: well-defined trajectories
in both space and time.
This new element constitutes the essence of what is known nowadays as
Bohmian mechanics \cite{bohm,holland-book}, a quantum mechanics based
on the hydrodynamical Eqs.~(\ref{eq14}) and (\ref{eq13}) plus the
particle equation of motion (\ref{eq16}).
Because of the relationship between particles and waves in Bohmian
mechanics, the initial momentum of particles is predetermined by
$\Psi({\bf r},0)$ and therefore it is not necessary to inquire about
its value.
Only initial positions are freely, randomly chosen, with the constraint
that their distribution is given by $\rho({\bf r},0)$.

As infers from Eq.~(\ref{eq14}), the difference with respect to
classical mechanics is that quantum trajectories evolve under the
action of both the external potential $V$ and the so-called
{\it quantum potential},
\be
 Q = - \frac{\hbar^2}{2m}\frac{\nabla^2 \rho^{1/2}}{\rho^{1/2}} ,
 \label{eq15}
\ee
which introduces into the quantum motion the context-dependence
and nonlocality necessary for the particles (distributed initially
according to $\rho({\bf r},0)$) to reproduce the patterns of standard
quantum mechanics when they are considered in a large statistical
number.
The quantum dynamics is thus ruled by a total effective potential
$V_{\rm eff}({\bf r},t) = V({\bf r}) + Q({\bf r},t)$.
An interesting property of the quantum potential can be found when
one computes the energy expected value,
\be
 \bar{E} = \langle \hat{H} \rangle
  = \langle \hat{T} \rangle + \langle \hat{V} \rangle ,
 \label{energy1}
\ee
which, in Bohmian mechanics, just consists of determining the ensemble
average energy.
The first term in (\ref{energy1}) is the expected value of the kinetic
energy, with $\hat{T} = - (\hbar^2/2m)\nabla^2$ denoting the kinetic
operator.
When the polar ansatz of the wave function is considered, we obtain
\be
 \langle \hat{T} \rangle = - \frac{\hbar^2}{2m}
  \int R \left[ \nabla^2 R
    - \frac{R}{\hbar^2} \left( \nabla S \right)^2
     \right] d{\bf r}
  + \frac{\hbar}{2im} \int \nabla (\rho \nabla S) d{\bf r} .
 \label{energy2}
\ee
From (\ref{eq13}), we have
\be
 \int \nabla {\bf J} d{\bf r} =
  - \int \frac{\partial \rho}{\partial t} \ \! d{\bf r} =
  - \frac{\partial}{\partial t} \int \rho \ \! d{\bf r} = 0 ,
 \label{energy3}
\ee
since the probability density conserves in the whole space.
In other words, if the system is closed, no probability can flow to or
from the region where this system is defined; this is the probabilistic
analog of the energy conservation principle.
Therefore, (\ref{energy2}) and (\ref{energy1}) become
\be
 \langle \hat{T} \rangle =
  \int \left( - \frac{\hbar^2}{2m} \frac{\nabla^2 R}{R} \right)
   \ \! \rho \ \! d{\bf r} +
  \int \frac{(\nabla S)^2}{2m} \ \! \rho \ \! d{\bf r}
  = \langle Q \rangle + \frac{\langle p^2 \rangle}{2m}
 \label{energy4}
\ee
and
\be
 \bar{E} = \bar{E_k} + \bar{Q} + \bar{V}
  = \bar{E_k} + \bar{V}_{\rm eff} ,
 \label{energy5}
\ee
respectively, where $\bar{E_k}$ is the average (expected value)
contribution from the kinetic energy associated with each particle
from the ensemble.
Thus, notice that, although $Q$ is usually assigned the role of
a quantum potential energy, since it is obtained from the kinetic
operator and contributes to the expected value of the kinetic energy,
one can also consider it as a quantum kinetic energy.
More importantly, putting aside such considerations, $Q$ is the only
responsible for making quantum dynamics so different from classical
ones due to its nonlocal nature ($\bar{E}_k$ and $\bar{V}$ only contain
local information) ---note that $Q$ contains information about the
whole quantum state and therefore supplies it to the particle at the
particular point where it is located.
From a practical or computational point of view, this explains why a
good representation of the momentum operator has to contain a wide
spectrum of momenta.
Furthermore, (\ref{energy5}) also indicates that classical-like
behaviors appear whenever the part of the energy associated with
the quantum potential becomes sufficiently small.

To conclude this section, we are going to discuss another interesting
property of quantum motion: in Bohmian mechanics two trajectories
cannot cross the same point at the same time in the configuration
space (note that, in classical mechanics, this only happens in the
phase-space).
In order to prove this, consider that two velocities are assigned
to the same point ${\bf r}$, $v_1({\bf r})$ and $v_2({\bf r})$, with
$v_1 \neq v_2$.
From (\ref{eq16}) we know that these velocities are directly related
to some wave functions $\psi_1({\bf r})$ and $\psi_2({\bf r})$,
respectively.
If both wave functions are solutions of the same Schr\"odinger
equation, the only possibility for them to be different at the
same space point $x$ is that their phases differ, as much, in a
time-dependent function (and/or space-independent constant), i.e.,
\be
 S_2({\bf r},t) = S_1({\bf r},t) + \phi(t) .
 \label{neweq}
\ee
Now, since the velocities are the gradient of these functions, it is
clear that $v_2$ has to be equal to $v_1$ necessarily if the wave
function does not vanish at ${\bf r}$.


\section{Simple examples of quantum interference with Gaussian
wave packets}
 \label{sec3}


\subsection{The free Gaussian wave packet}
 \label{sec3.1}

For the sake of simplicity, we are going to consider one-dimensional
Gaussian wave packets, although the results presented in this work
can be generalized to more dimensions and other types of wave packets
straightforwardly.
The evolution of a free Gaussian wave packet can be described by
\be
 \Psi(x,t) = A_t \ \! e^{- (x-x_t)^2/4\tilde{\sigma}_t\sigma_0
  + i p (x - x_t)/\hbar + i E t/\hbar} ,
 \label{eqn2}
\ee
where $A_t = (2\pi\tilde{\sigma}_t^2)^{-1/4}$ and the complex
time-dependent spreading is
\be
 \tilde{\sigma}_t =
  \sigma_0 \left( 1 + \frac{i\hbar t}{2m\sigma_0^2} \right) .
 \label{eqn3}
\ee
From (\ref{eqn3}), the spreading of this wave packet at time $t$ is
\be
 \sigma_t = |\tilde{\sigma}_t| =
  \sigma_0 \sqrt{1 + \left( \frac{\hbar t}{2m\sigma_0^2} \right)^2 } .
 \label{eqn4}
\ee
Due to the free motion, $x_t = x_0 + v_p t$ ($v_p = p/m$ is the
propagation velocity) and $E = p^2/2m$, i.e., the centroid of the
wave packet moves along a classical rectilinear trajectory.
This does not mean, however, that the expected (or average) value
of the wave-packet energy is equal to $E$, which, according to
Eq.~(\ref{energy5}), results
\be
 \bar{E} = \frac{p^2}{2m} + \frac{\hbar^2}{8m\sigma_0^2} .
 \label{energy5b}
\ee
In (\ref{energy5b}) we thus observe two contributions.
The first one is associated with the propagation or translation of
the wave packet, while the latter is related to its spreading and
has, therefore, a purely quantum-mechanical origin.
From now on, we will denote them as $E_p$ and $E_s$, respectively,
noting that none of them depends on time.
Expressing $E_s$ in terms of an {\it effective} momentum $p_s$
(i.e., $E_s = p_s^2/2m$), we find
\be
 p_s = \frac{\hbar}{2\sigma_0} ,
 \label{heisen}
\ee
which resembles Heisenberg's uncertainty relation.
Note that, in the case of {\it non-dispersive} Airy wave packets
\cite{airy}, we will find $p_s = 0$.

With the previous definition, the wave packet dynamics can be referred
to two well-defined velocities: $v_p$ and $v_s$, whose relationship is
shown to play an important role in the type of effects that one can
observe when dealing with wave packet superpositions.
To better understand this statement, consider Eq.~(\ref{eqn4}).
Defining the timescale
\be
 \tau = 2m\sigma_0^2/\hbar ,
 \label{tau}
\ee
we find that, if $t \ll \tau$, the width of the wave
packet remains basically constant with time, $\sigma_t \approx
\sigma_0$ (i.e., for practical purposes, it is roughly time-independent
up to time $t$).
On the contrary, if $t \gg \tau$, the width of the wave packet
increases linearly with time ($\sigma_t \approx \hbar t/2m\sigma_0$).
Of course, in between there is a smooth transition; from (\ref{eqn4})
it is shown that the progressive increase of $\sigma_t$ describes a
hyperbola when this magnitude is plotted {\it vs} time.
These are ``dynamical'' relationships in the sense that the dynamical
behavior of the wave packet is known comparing the actual time $t$
with the effective time $\tau$.
When we are dealing with general wave packets, {\it a priori} we do not
have an effective time $\tau$ to compare with.
This inconvenience can be avoided if we consider the interpretive
framework provided by $v_p$ and $v_s$, which can be estimated from
the initial state through Eq.~(\ref{energy4}).
Within this framework, we can find a way to know which process
---spreading or translational motion--- is going to determine the
wave packet dynamics, as follows.
Expression (\ref{eqn4}) can be rewritten in terms of $v_s$ as
\be
 \sigma_t = |\tilde{\sigma}_t| =
  \sigma_0 \sqrt{1 + \left( \frac{v_s t}{\sigma_0} \right)^2 } .
 \label{eqn4bis}
\ee
Now, consider that $t$ is the time that takes the wave-packet centroid
to cover a distance $d = v_p t \approx \sigma_0$.
Introducing this result into (\ref{eqn4bis}), we obtain
\be
 \sigma_t = |\tilde{\sigma}_t| =
  \sigma_0 \sqrt{1 + \left( \frac{v_s}{v_p} \right)^2 } .
 \label{eqn4bbis}
\ee
This time-independent expression shows that, only using information
about the initial preparation of the wave packet, we can obtain
information on its subsequent dynamical behavior.
Thus, $t \ll \tau$ is equivalent to having an initial wave packet
prepared with $v_s \ll v_p$: the translational motion will be much
faster than the spreading of the wave packet.
On the other hand, $t \gg \tau$ is equivalent to $v_s \gg v_p$: the
wave packet spreads very rapidly in comparison with its advance along
$x$.
A nice illustration of the ``competition'' between $v_p$ and $v_s$ can
be found in \cite{sanz-cpl}, where (for $|x_0| \approx \sigma_0$) it
is shown that: (1) if $v_p$ is dominant, the asymptotic motion is
basically a classical like motion (no spreading in comparison with
the distances covered by particles), while (2) if $v_s$ is dominant,
although the motion is classical-like (i.e., there is an asymptotic
constant velocity, $v_s$), it has a purely quantum-mechanical origin
and therefore eventual effects produced within this regime will also
be purely quantum-mechanical.


\subsection{Dynamics of coherent wave-packet superpositions}
 \label{sec3.2}

Depending on whether the interference is localized in time within a
certain space region or it remains stationary after some time, we can
distinguish two types of processes or experiments: collision-like
(e.g., BEC interferometry) and diffraction-like (e.g., the double-slit
experiment), respectively.
These behaviors are associated with the ratio between $v_p$ and $v_s$:
collision-like experiments are characterized by $v_p \gg v_s$, while
diffraction-like experiments by $v_p \ll v_s$.
Now, consider a general coherent superposition
\be
 \Psi = c_1 \psi_1 + c_2 \psi_2 ,
 \label{sup1}
\ee
where the partial waves $\psi_i$ are normalized Gaussian wave packets,
as in (\ref{eqn2}), which propagate with opposite velocities $v_p$ and
where the centers are initially far enough in order to minimize the
overlapping, i.e., $\rho_1({\bf r},0) \rho_2({\bf r},0) \approx 0$.
As shown below, both the $v_p/v_s$ ratio and the weighting factors
$c_1$ and $c_2$ are going to influence the topology of the quantum
trajectories.
Since the wave packets are well apart initially, we find that the
weighting factors satisfy the relation $|c_1|^2 + |c_2|^2 = 1$, which
allows us to reexpress (\ref{sup1}) more conveniently as
\be
 \Psi = c_1 \left( \psi_1 + \sqrt{\alpha} \psi_2 \right) ,
 \label{sup2}
\ee
where $\alpha = (c_2/c_1)^2$.
From (\ref{sup2}), we find that the probability density and the quantum
current density read as
\ba
 \rho & = & c_1^2 \left[ \rho_1 + \alpha \rho_2
  + 2 \sqrt{\alpha} \sqrt{\rho_1 \rho_2} \cos \varphi \right] ,
 \label{sup4} \\
 {\bf J} & = & \frac{c_1^2}{m} \ \!
  \Big[ \rho_1 \nabla S_1 + \alpha \rho_2 \nabla S_2
  + \sqrt{\alpha} \sqrt{\rho_1 \rho_2} \nabla (S_1 + S_2) \cos \varphi
 \nonumber \\
  & & + \hbar \sqrt{\alpha} (\rho_1^{1/2} \nabla \rho_2^{1/2}
    - \rho_2^{1/2} \nabla \rho_1^{1/2}) \sin \varphi \Big] ,
 \label{sup5}
\ea
respectively.
Also, dividing Eq.~(\ref{sup5}) by (\ref{sup4}), as in (\ref{eq16}),
we can obtain the associate quantum trajectories from the equation of
motion
\ba
 \dot{\bf r} & = & \frac{1}{m}
  \frac{\rho_1 \nabla S_1 + \alpha \rho_2 \nabla S_2 +
   \sqrt{\alpha} \sqrt{\rho_1 \rho_2} \nabla (S_1 + S_2) \cos \varphi}
  {\rho_1 + \alpha \rho_2 +
   2 \sqrt{\alpha} \sqrt{\rho_1 \rho_2} \cos \varphi}
 \nonumber \\
  & & + \sqrt{\alpha} \ \! \frac{\hbar}{m}
  \frac{(\rho_1^{1/2} \nabla \rho_2^{1/2}
    - \rho_2^{1/2} \nabla \rho_1^{1/2}) \sin \varphi}
  {\rho_1 + \alpha \rho_2 +
   2 \sqrt{\alpha} \sqrt{\rho_1 \rho_2} \cos \varphi} .
 \label{sup7}
\ea
As can be noticed, in this expression there are two well-defined
contributions, which are related to the effects caused by the
interchange of the wave packets (or the associate partial fluxes) on
the particle motion (specifically, on the topology displayed by the
corresponding quantum trajectories) after the interference process;
note that the first contribution is even after interchanging only the
modulus or only the phase of the wave packets, while the second one
changes its sign with these operations.
From the terms that appear in each contribution, it is apparent that
the first contribution is associated with the evolution of each
individual flux as well as with their combination.
Thus, it provides information about both the asymptotic behavior of
the quantum trajectories and also about the interference process
(whenever the condition $\rho_1({\bf r},t) \rho_2({\bf r},t) \approx 0$
is not satisfied).
On the other hand, the second contribution describes interference
effects connected with the asymmetries or differences of the wave
packets.
For instance, their contribution is going to vanish if they are
identical and coincide at $x=0$ although their overlapping is nonzero.

\begin{figure}
 \begin{center}
 \epsfxsize=16cm {\epsfbox{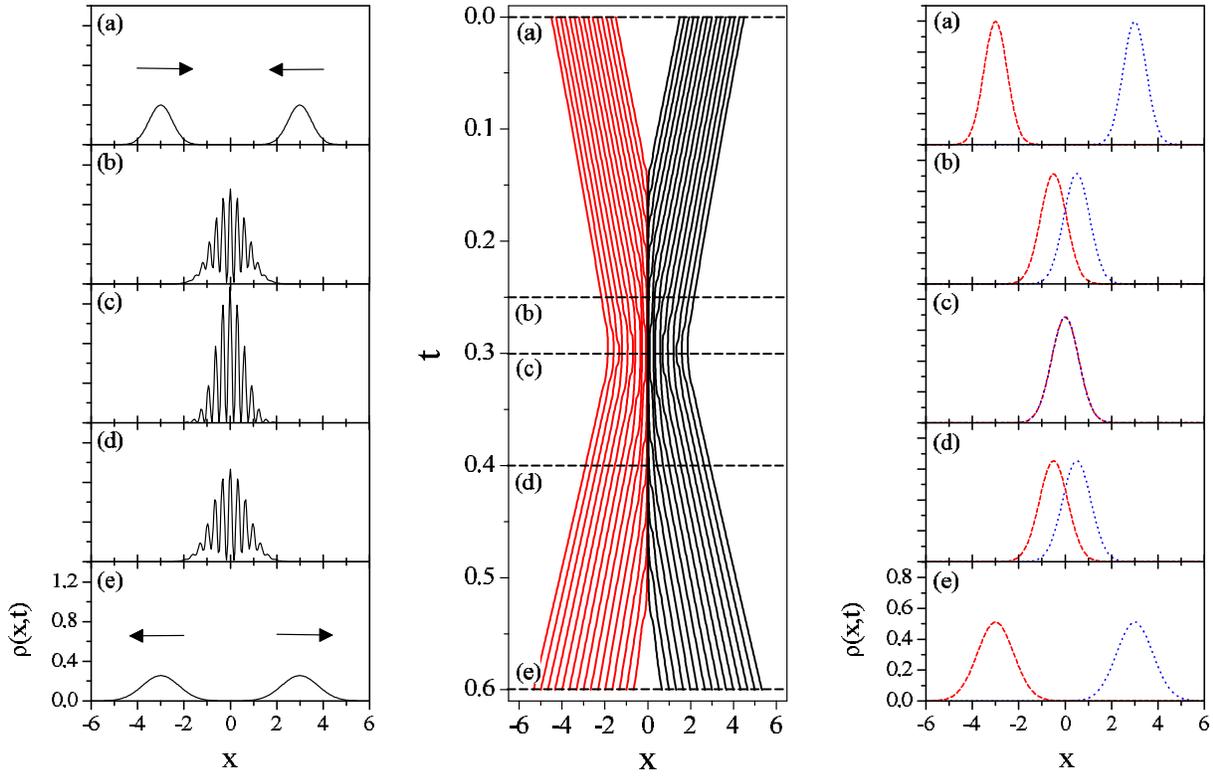}}
 \caption{\label{fig2}
  Left: From (a) to (d), snapshots illustrating the time-evolution of
  two colliding Gaussian wave packets, with $v_p = 10$ and $v_s = 1$.
  Center: The same process as in the left column, by visualized in
  terms of quantum trajectories.
  The perpendicular dashed lines mark cuts in time corresponding with
  the same labels as in the figures of the left column.
  Right: Interpretation of the process in the left column according to
  the quantum trajectories shown in the central column.}
 \end{center}
\end{figure}

Let us first consider the collision-like case.
For the sake of simplicity and without loss of generality, we will
assume $\alpha = 1$, with both wave packets being identical and
propagating in opposite directions at the same speed.
Also, for simplicity in the discussion, we will refer to the regions
where $\psi_1$ and $\psi_2$ are initially placed as I and II,
respectively.
In standard quantum mechanics, a physical reality is assigned to the
superposition principle, which in this case means that, after the wave
packets have maximally interfered at $t^{\rm int}_{\rm max}$ (panel (c)
in the left column of Fig.~\ref{fig2}), $\psi_1$ moves to region II and
$\psi_2$ to region I.
As mentioned above, the wave packets (or, more specifically, their
associate probability densities) represent the statistical behavior
of a swarm of identical, noninteracting quantum particles distributed
accordingly.
Therefore, within the standard view arising from the superposition
principle, one would expect to observe crossings between trajectories
in a certain time range around $t^{\rm int}_{\rm max}$.
However, this is not the behavior displayed by the Bohmian trajectories
displayed in the central panel of Fig.~\ref{fig2}, which avoid such
crossings during the time range where interference effects are
important.
Note that quantum-mechanical statistics are characterized by keeping
the coherence and transmitting it to the corresponding quantum
dynamics, as infers from Eq.~(\ref{sup7}).
Thus, the interference process has to be interpreted in a different way
(than the standard quantum-mechanical one) when it is analyzed from a
quantum trajectory perspective.
As inferred from Eq.~(\ref{sup7}), for identical (but
counter-propagating) wave packets, the velocity field vanishes
along $x=0$ at any time.
This means that there cannot be any probability density (or particle)
flux between regions I and II at any time.
Therefore, trajectories starting in one of these regions will never
cross to the other one and vice versa.
Hence the final outgoing wave packets in panel (e) in the left column
of Fig.~\ref{fig2} necessarily describe exactly the same swarms of
particles associated with the wave packets that we had in their
respective regions initially (see panel (a) of the same figure).
The whole process can be then understood as a sort of bouncing motion
of the wave packets once they have reached the intermediate position
between them, as schematically represented in the right column of
Fig.~\ref{fig2}.

\begin{figure}
 \begin{center}
 \epsfxsize=7cm {\epsfbox{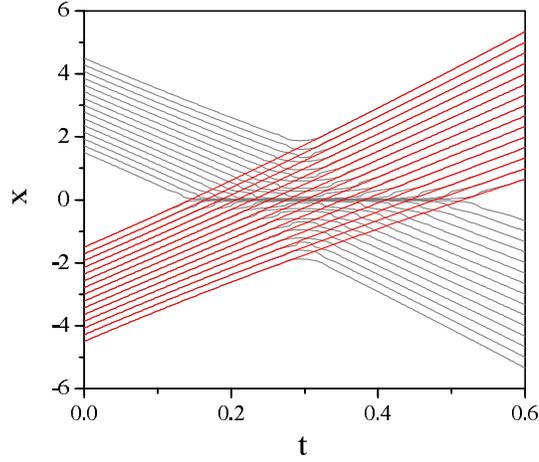}}
 \caption{\label{fig3}
  Bohmian trajectories associated with a Gaussian wave-packet
  superposition (grey) and a single Gaussian wave packet (red).
  As shown, the presence of the other wave packet makes the dynamics
  to avoid the crossing in the central part; the asymptotic part
  remains being the same as without presence of other wave packet.
  Here $v_p = 10$ and $v_s = 1$.}
 \end{center}
\end{figure}

But, if the two swarms of trajectories do not cross each other, why
the swarm associated initially with one of the wave packets behaves
asymptotically as associated with the other one, as seen in
Fig.~\ref{fig3}?
According to the standard description based on the superposition
principle, the wave packets cross.
This can be understood as a transfer or interchange of probabilities
from region I to region II and vice versa.
On the other hand, from a dynamical (quantum trajectory) viewpoint,
this also means that the sign of the associate velocity field will
change after the collision.
That is, before the collision its sign points onwards (towards $x=0$)
and after the collision it points outwards (diverging from $x=0$)
---at $t^{\rm int}_{\rm max}$ it does not point anywhere, but remains
{\it steady}.
Thus, like in a particle-particle elastic scattering
process particles exchange their momenta, here the swarms of particles
will exchange their probability distributions ``elastically''.
This is nicely illustrated in Fig.~\ref{fig3}: after the collision,
the two swarms of trajectories (grey lines) bounce backwards and follow
the paths that would be pursued by non deflected particles (red lines).
We can describe this process analytically as follows.
Initially, depending on the region where the trajectories are launched
from, they are described approximately by
\be
 \dot{\bf r}_{\rm I} \approx \nabla S_1/m  \qquad {\rm or} \qquad
 \dot{\bf r}_{\rm II} \approx \nabla S_2/m .
\ee
Now, asymptotically (for $t \gg t^{\rm int}_{\rm max}$),
Eq.~(\ref{sup7}) can be expressed as
\be
 \dot{\bf r} \approx \frac{1}{m}
  \frac{\rho_1 \nabla S_1 + \rho_2 \nabla S_2}{\rho_1 + \rho_2} ,
 \label{sup7b}
\ee
where we make use of the approximation $\rho_1({\bf r},t)
\rho_2({\bf r},t) \approx 0$.
Note that this approximation also means that the total probability
density, $\rho = \rho_1 + \rho_2$, is nonzero only on the space
regions covered by either $\rho_1$ or $\rho_2$.
Specifically, in region I we will have $\rho \approx \rho_2$ and in
region II, $\rho \approx \rho_1$.
When this result is introduced into Eq.~(\ref{sup7b}), we finally
obtain
\be
 \dot{\bf r}_{\rm I} \approx \nabla S_2/m  \qquad {\rm and} \qquad
 \dot{\bf r}_{\rm II} \approx \nabla S_1/m ,
\ee
which reproduce the dynamics observed asymptotically.
This trajectory picture thus provides us with a totally different
interpretation of wave packet interference with respect to the standard
one: although probability distributions transfer, particles remain
always within the domains defined by their corresponding initial
distributions (which have been here as regions I and II).
The {\it non-crossing} property of Bohmian mechanics (see
section~\ref{sec2}) thus makes apparent the constraint that exists for
the quantum probability flux, which goes beyond the separability of
fluxes implicit in the superposition principle.

In the diffraction-like case, the spreading is faster than the
propagation, this being the reason why we observe the well-known
quantum trajectories of a typical two-slit experiment in the
{\it Fraunhofer region} \cite{sanz-jpcm,sanz-prbI}.
It is relatively simple to show \cite{sanz-talbot} that, in this case,
the asymptotic solutions of Eq.~(\ref{sup7}) are
\be
 x(t) \approx
 2\pi n \ \! \frac{\sigma_0}{x_0} \left( v_s t \right) ,
 \label{approx5}
\ee
with $n = 0, \pm 1, \pm 2, \ldots$
That is, there are bunches of quantum trajectories whose slopes are
quantized quantities (through $n$) proportional to $v_s$.
This means that, when Eq.~(\ref{sup7}) is integrated exactly, one will
observe quantized bunches of trajectories which, on average, are
distributed around the value given by Eq.~(\ref{approx5}).


\subsection{Asymmetric coherent superpositions}
 \label{sec3.3}

In the previous subsection we have studied interference processes with
identical Gaussian wave packets propagating symmetrically with respect
to $x=0$.
However, what happens in other more general situations where neither
the wave packets are identical nor their weights?
Although one could think of many different possibilities, it is enough
to consider three cases in order to already obtain an insight on the
related physics.
The criterion followed to classify these cases is based on which
property of the wave packets is varied (with respect to the symmetric
case described in section~\ref{sec3.2}) or considered at a time:
\begin{itemize}
 \item Case A: Different modulus of the initial average momentum
 ($|p_{01}| \ne |p_{02}|$).
 \item Case B: Different initial spreading
 ($\sigma_{01} \ne \sigma_{02}$).
 \item Case C: Different weights ($c_1 \ne c_2$ or, equivalently,
 $\alpha \ne 1$).
\end{itemize}
Since the collision-like case is simpler to understand than the
diffraction-like one, below we will assume $v_p \gg v_s$, although
generalizing to any $v_p/v_s$ ratio is straightforward.
Moreover, as before, the initial distance between both wave packets is
such that the initial overlapping is negligible.

\begin{figure}
 \begin{center}
 \epsfxsize=14cm {\epsfbox{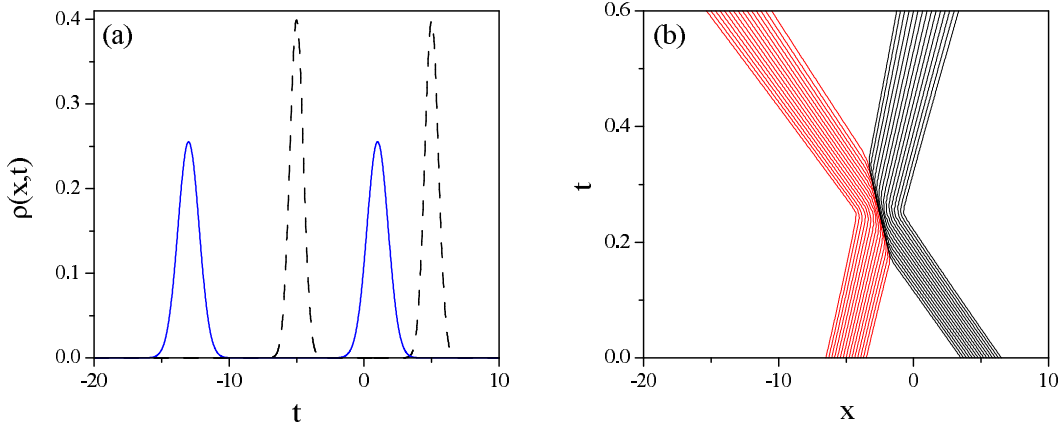}}
 \caption{\label{fig5}
  (a) Probability density at $t=0$ (black-dashed line) and $t=0.6$
  (blue-solid line) for a wave packet superposition with $p_{01} = 10$
  and $p_{02} = -30$.
  The value of the other relevant parameters are $\sigma_{01} =
  \sigma_{02} = 0.5$ and $\alpha = 1$.
  (b) Quantum trajectories associated with the dynamics described by
  the situation displayed in panel (a).}
 \end{center}
\end{figure}

Case A is represented in Fig.~\ref{fig5}, where we can observe that
varying the moduli of the initial average momenta only produces an
asymmetric shift of the final positions of the wave packets with
respect to $x=0$ (see panel (a)), which is a result of the distortion
of the boundary between regions I and II (the final wave packets are
not symmetrically distributed with respect to $x=0$).
This boundary evolves in time at the same constant velocity than
the corresponding expected value of the velocity for the total
superposition,
\be
 \bar{v}_{\rm A} = \frac{\bar{p}_{\rm A}}{m}
  = \frac{\langle \Psi_{\rm A}^* | \hat{p} | \Psi_{\rm A} \rangle}{m}
  = \frac{p_{01} + p_{02}}{2m} ,
 \label{momav}
\ee
because both wave packets are identical and normalized (see
Eq.~(\ref{eqn2})).
In particular, since $p_{01} = 10$ and $p_{02} = -30$, we have
$\bar{v}_{\rm A} = -10$.
Integrating the equation of motion $\dot{\bar{x}}_{\rm A} =
\bar{v}_{\rm A}$ (according to Ehrenfest's theorem \cite{schiff}),
we obtain
\be
 \bar{x}_{\rm A}(t) = \bar{x}_0 + \bar{v}_{\rm A} t ,
 \label{momavAnc}
\ee
with $\bar{x}_0 = \langle \Psi^* | \hat{x} | \Psi \rangle = (x_{01} +
x_{02})/2$.
Equation~(\ref{momavAnc}) describes the time-evolution of the boundary
between regions I and II in this case ---note that for $p_{\rm 02} =
- p_{01}$, we recover again the time-independent boundary found for
identical wave packets, studied in the previous section.
As expected, this boundary also defines the non-crossing line for the
associate quantum trajectories, which forbids the trajectory transfer
from one region to the other one and vice versa, as seen in
Fig.~\ref{fig5}(b).
Because of this property, we find that the two wave packets behave like
two classical particles undergoing an elastic scattering: there is only
transfer (indeed, exchange) of momentum during the scattering process,
but the net balance of probability is zero since no trajectories are
transferred.
Moreover, note that indeed this effect will not be noticeable unless
one looks at the quantum trajectories ---as seen in panel (a), the
evolution of the wave packets does not provide any clue on it.

\begin{figure}
 \begin{center}
 \epsfxsize=14cm {\epsfbox{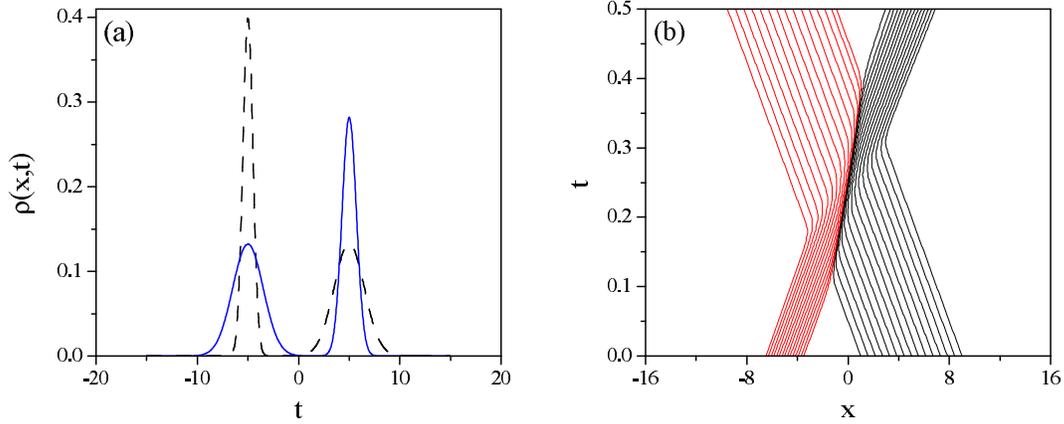}}
 \caption{\label{fig6}
  (a) Probability density at $t=0$ (black-dashed line) and $t=0.8$
  (blue-solid line) for a wave packet superposition with $\sigma_{01}
  = 0.5$ and $\sigma_{02} = 1.5$.
  The value of the other relevant parameters are $|p_{01}| = |p_{02}|
  = 20$ and $\alpha = 1$.
  (b) Quantum trajectories associated with the dynamics described by
  the situation displayed in panel (a).}
 \end{center}
\end{figure}

In case B, since both velocities are equal in modulus, after
interference the centers of the outgoing wave packets occupy symmetric
positions with respect to $x=0$, as seen in Fig.~\ref{fig6}(a) (only
the different width disturbs the full symmetry).
However, as in case A, the lack of total symmetry also causes the
distortion of the boundary between regions I and II (see panel (b)),
with a similar behavior regarding the non-crossing (or non-transfer)
property.
Now, since the modulus of both momenta are the same,
$\langle \Psi^* | \hat{p} | \Psi \rangle = 0$ and therefore we cannot
appeal to the same argument as before to explain this distortion
effect.
However, there is still a sort of effective ``internal'' momentum which
depends on the ration between the widths of the wave packets.
Assuming that the spreading is basically constant for the time we have
propagated the trajectories, the corresponding effective velocity is
given by
\be
 \bar{v}_{\rm B}
  = \frac{\sigma_{01}^{-1} \ \! v_{01} + \sigma_{02}^{-1} \ \! v_{02}}
    {\sigma_{01}^{-1} + \sigma_{02}^{-1}}
  = \left( \frac{\sigma_{02} - \sigma_{01}}{\sigma_{01} + \sigma_{02}}
     \right) v_0 ,
 \label{momavB}
\ee
with $v_{01} = v_0 = - v_{02}$ ($v_0 > 0$); since the width of the
wave packets varies in time, a slight time-dependence can be expected
in $\bar{v}_{\rm B}$, although in our case, as can be noticed from
Fig.~\ref{fig6}(b), it can be neglected for practical purposes.
Substituting the numerical values used in the propagation into
(\ref{momavB}), we find $\bar{v}_{\rm B} = 5$.
The boundary or non-crossing line will be then given by
\be
 \bar{x}_{\rm B}(t) = \bar{x}_0 + \bar{v}_{\rm B} t ,
 \label{momavAncb}
\ee
where
\be
 \bar{x}_0
  = \frac{\sigma_{01}^{-1} \ \! x_{01} + \sigma_{02}^{-1} \ \! x_{02}}
    {\sigma_{01}^{-1} \ \! + \sigma_{02}^{-1}}
  = - \left( \frac{\sigma_{02} - \sigma_{01}}
      {\sigma_{01} + \sigma_{02}} \right) x_0 ,
 \label{momavBb}
\ee
with $x_{02} = - x_{01} = x_0$ ($x_0 > 0$) ---note that, otherwise,
$\bar{x}_0=0$, as infers from Fig.~\ref{fig6}(a), since the larger
width of one of the wave packets balances the effect of the larger
height of the other one.
Despite this internal redistribution or balance of momentum, it is
clear that we find again, as in case A, an elastic collision-like
behavior.

The effect of an effective internal momentum could be explained by
considering that, in this case, the relative size of the wave packets
acts like a sort of {\it quantum inertia} or effective mass.
Consider the trajectories associated with the wave packet with smaller
value of $\sigma_0$ (black trajectories in Fig.~\ref{fig6}(b)).
As can be noticed, it takes approximately 0.2 time units the whole
swarm to leave the scattering or collision region (along the
non-crossing boundary) ---or, equivalently, to revert the sign of the
momentum of all the trajectories constituting the swarm and get their
final asymptotic momenta.
On the other hand, the trajectories associated with the wave packet
with larger $\sigma_0$ (red trajectories in Fig.~\ref{fig6}(b)) revert
their momenta much faster, in about 0.1 time units ---of course, we are
not considering here the time that trajectories keep moving along the
boundary, since the total ``interaction'' time has to be the same for
both swarms of trajectories.
Thus, we find that the larger the spreading momentum, the larger the
quantum inertia of the swarm of particles to change the propagation
momentum and, therefore, to reach the final state.

\begin{figure}
 \begin{center}
 \epsfxsize=7cm {\epsfbox{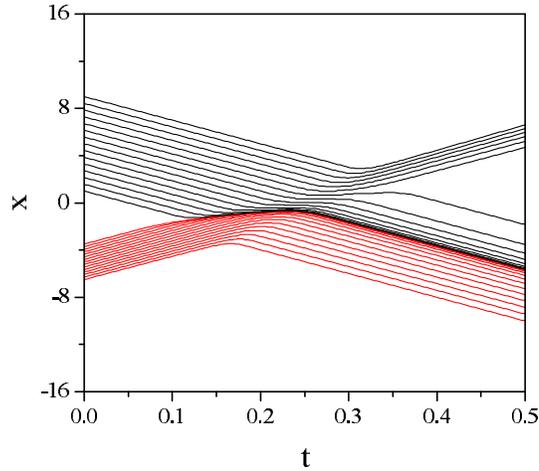}}
 \caption{\label{fig7}
  Bohmian trajectories associated with a Gaussian wave-packet
  superposition with $\sigma_{01} = 0.5$ and $\sigma_{02} = 1.5$.
  The value of the other relevant parameters are $p_{01} = - p_{02}
  = 20$ and $\alpha = 1$.}
 \end{center}
\end{figure}

We would also like to note another interesting property associated with
case B.
Consider that both wave packets lack their corresponding normalizing
factor $A_t$ when introduced in the superposition.
If we then compute the expected value of the momentum, for instance,
we obtain
\be
 \langle \hat{p} \rangle_{\rm B}
  = \frac{\langle \Psi_{\rm B}^* | \hat{p} | \Psi_{\rm B} \rangle}
     {\langle \Psi_{\rm B}^* | \Psi_{\rm B} \rangle}
  = \frac{\sigma_{01} \ \! v_{01} + \sigma_{02} \ \! v_{02}}
    {\sigma_{01} + \sigma_{02}}
  = \left( \frac{\sigma_{01} - \sigma_{02}}{\sigma_{01} + \sigma_{02}}
     \right) p_0 ,
 \label{momavBbis}
\ee
i.e., there should be a certain ``drift'' towards region I, such as in
case A ---and the same holds if we compute instead the expected value
of the position.
Note that in the previous case the normalization of each Gaussian wave
packet produces a balance: the probability with which each wave packet
contributes to the superposition is the same ($c_1^2 = c_2^2 = 1/2$)
because the width of one compensates the height of the other, as
explained above.
Therefore, the expected value of both position and momentum have to be
zero.
However, this compensation does not happen now: both wave packets have
the same height although their widths differ, thus contributing with
different probabilities $P$ to the superposition,
\be
 P_1 = \frac{\sigma_{01}}{\sigma_{01} + \sigma_{02}} \qquad {\rm and}
  \qquad P_2 = \frac{\sigma_{02}}{\sigma_{01} + \sigma_{02}} ,
 \label{probs}
\ee
which produce the results observed in Fig.~\ref{fig7} (again, we
assume that $\sigma_t \approx \sigma_0$ for the time considered).
However, by inspecting (\ref{momavBbis}), we note that if we add the
averaged momentum $\bar{p}_{\rm B} = m\bar{v}_{\rm B}$, the total
average momentum vanishes.
Somehow the averaging defined by (\ref{momavBbis}) acts as in classical
mechanics, when a certain magnitude (e.g., the position or the
momentum) is computed with respect to the center of mass of a system.
Here, $\bar{p}_{\rm B}$ is the magnitude necessary to reset the
superposition of non normalized Gaussian wave packets to a certain
``center of spreading''.
On the other hand, it is also important to stress the fact that, in
this case, the clear boundary between the swarms of trajectories
associated with each initial wave packet disappears.
Now, although there is still a boundary, it does not prevent for the
transfer of trajectories from one region to the other, as before.
This effect, similar to consider inelastic scattering in classical
mechanics, arises as a consequence of having wave packets with
different probabilities, which we analyze below.

\begin{figure}
 \begin{center}
 \epsfxsize=14cm {\epsfbox{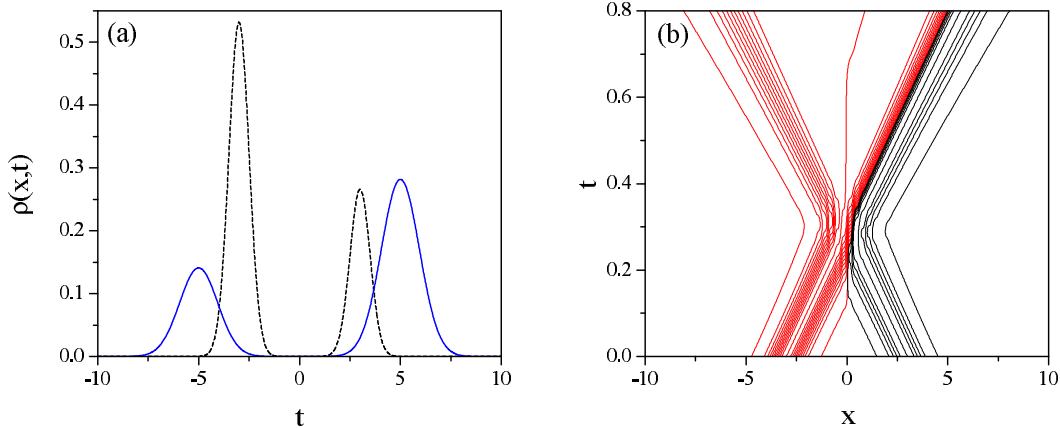}}
 \caption{\label{fig8}
  (a) Probability density at $t=0$ (black-dashed line) and $t=0.8$
  (red-solid line) for a wave packet superposition with $\alpha=0.5$
  (i.e., $c_1^2=2/3$ and $c_2^2=1/3$).
  The value of the other relevant parameters are $p_{01}=-p_{02}=10$
  and $\sigma_{01} = \sigma_{02} = 0.5$.
  (b) Quantum trajectories associated with the dynamics described by
  the situation displayed in panel (a).
  The initial conditions for the quantum trajectories have been
  assigned by considering the different weights associated with
  each wave packet.}
 \end{center}
\end{figure}

As seen above, unless there is an asymmetry in the probabilities
carried by each partial wave in the superposition, there is always a
well-defined boundary or non-crossing line between regions I and II.
However, in the second example of case B, we have observed that it
is enough an asymmetry in the probability distribution to break
immediately the non-crossing line.
Instead of using non normalized Gaussian wave packets, let us consider
case C, which is equivalent although the asymmetry is caused by
$\alpha \ne 1$ instead of the wave-packet normalization.
As we have seen in the two previous cases, provided both wave packets
have the same weight (or, at least, both contribute equally to the
superposition), the corresponding quantum trajectories, even if
they are evenly distributed (equidistant) along a certain distance
initially, they are going to give a good account of the whole
dynamics.
However, if the weights change, the same does not hold anymore.
Somehow equal weights (or equal probabilities) means the division of
the coordinate space into two identical regions, each one influenced
only by the corresponding wave packet.
In other words, there are always two well-defined swarms of
trajectories, each one associated with one of the initial wave
packets.
When the weights change, apparently we have something similar to what
we have been observing until now: the wave packets exchange their
positions (see Fig.~\ref{fig8}(a)).
However, when we look at the associate quantum trajectories, we realize
that there is transfer or flow of trajectories from one of the initial
swarms to the other one.
This transfer takes place from the swarm with larger weight to the
lesser one, thus distorting importantly the boundary between regions
I and II, as seen in Fig.~\ref{fig8}(b), where this boundary lies
somewhere between the two trajectories represented in blue (the
trajectories that are closer to $x=0$ in each swarm).
However, note that this does not imply that the number of trajectories
varies in each region asymptotically, but only the number of them
belonging to one or the other wave packet.
Thus, if initially we have $N_1 \propto c_1^2$ trajectories associated
with $\psi_1$ and $N_2 \propto c_2^2$ with $\psi_2$, asymptotically we
will observe $N'_1 = (1 - \alpha) N_1 \propto c_2^2$ and
$N'_2 = N_2 + \alpha N_1 \propto c_1^2$ due to the trajectory transfer.
Unlike the two previous cases discussed above, this process can be
then compared with inelastic scattering, where, after collision, not
only the probability fluxes but also the number of particles changes.
Accordingly, it is also important to mention that, due to the
trajectory transfer, representations with evenly distributed
trajectories are not going to provide a good picture of the problem
dynamics, as infers from Fig.~\ref{fig7}.
Rather, we can proceed in two different ways.
The obvious procedure is to consider initial positions distributed
according to the corresponding (initial) probability densities.
This procedure carries a difficulty: the number of trajectories needed
to have a good representation of the dynamics may increase enormously
depending on the relative weights.
The second procedure is to consider evenly spaced values of the
probability density and allocate at such space points the initial
positions of the trajectories, as we have done to construct
Fig.~\ref{fig8}(b).
In this case, although the trajectories will not accumulate exactly
along the regions with larger values of the probability density, this
construction has the advantage that we can follow the transport of
equi-spaced probabilities along each particular trajectory.
To make more apparent the difference in the relative number of
trajectories (probability density) associated with each initial wave
packets, we have considered $N_1 = 23$ and $N_2 = 11$ (i.e.,
$N_2/N_1 \sim 0.48 \approx c_2^2/c_1^2 = 0.5$).
After scattering, the numbers that we have are $N'_1=11$ and $N'_2=23$,
which are in the expected ratio ${c'}_2^2/{c'}_1^2 = c_1^2/c_2^2
= 2$.

As said at the beginning of this subsection, here we have only considered
the collision-like case.
The same kind of results are expected for the analogous
diffraction-like cases, with the addition that they manifest as a loss
of fringe visibility.
Thus, for cases A and B, one could appreciate a well define
non-crossing line and a loss of fringe visibility due to the divergent
velocities, $v_p$ and $v_s$, respectively.
And, for case C (or the second example in case B), the loss of fringe
visibility would be caused by the transfer of trajectories, which
would also lead to the distortion of the non-crossing line.


\subsection{Coherent superpositions, potential barriers and resonances}
 \label{sec3.4}

\begin{figure}
 \begin{center}
 \epsfxsize=14cm {\epsfbox{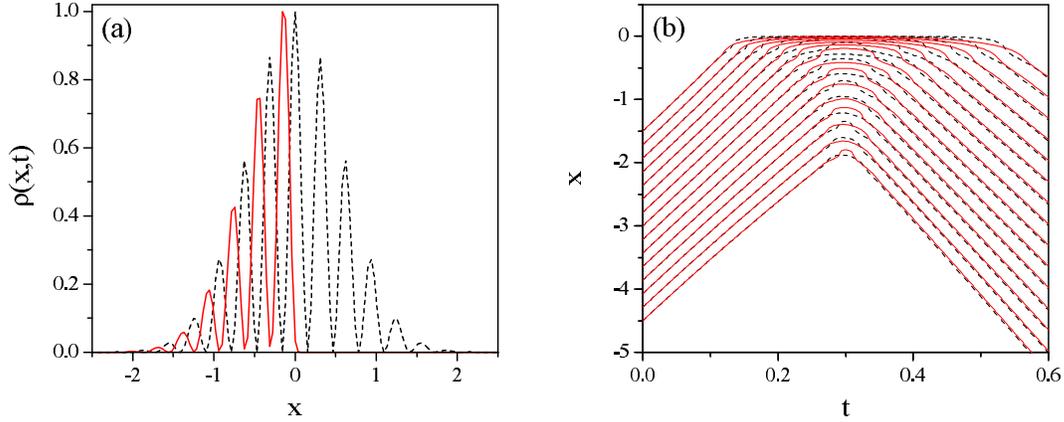}}
 \caption{\label{fig9}
  (a) Probability density at $t=0.3$ for the collision of a Gaussian
  wave packet off: an external impenetrable wall (red-solid line) and
  another (identical) Gaussian wave packet (black-dashed line).
  To compare, the maxima of both probability densities are normalized
  to unity.
  (b) Bohmian trajectories illustrating the dynamics associated with
  the two cases displayed in part (a).
  Here, $v_p = 10$ and $v_s = 1$.}
 \end{center}
\end{figure}

Finally, here we are going to analyze an interesting effect associated
with the non-crossing property described in the previous sections.
Consider a Gaussian wave packet scattered off an impenetrable potential
wall (for simplicity, we will assume $v_p > v_s$ for now).
After some time the wave packet will collide with the wall and then
part of it will bounce backwards.
The interference of the forward ($f$) and backward ($b$) wave packets
will lead to a fringe-like pattern similar to those observed in the
previous sections (see red-solid line in Fig.~\ref{fig9}(a)), with
also a time, $t^{\rm int}_{\rm max}$, for which the interference
fringes are maximally resolved.
Putting aside the initial Gaussian shape of the wave packet (and,
therefore, the effects associated with $v_s$), if this process is
represented as
\be
 \Psi = \psi_f + \psi_b
  \sim e^{imv_p x/\hbar} + e^{-imv_p x/\hbar}
 \label{inout}
\ee
at $t^{\rm int}_{\rm max}$, we obtain $\rho(x)\sim\cos^2(mv_px/\hbar)$.
The distance between two consecutive minima is then $w_0=\pi\hbar/mv_p$,
which turns out to be the same distance between two consecutive
minima in the two wave-packet interference process (see black-dashed
line in Fig.~\ref{fig9}(a)).
That is, although each process has a different physical origin (barrier
scattering {\it vs} wave packet collision), the effect is similar
---there is a certain shift in the position of the corresponding
maxima ($\sim \pi/2$), which arises from the fact that, in the case of
barrier scattering, the impenetrable wall forces the wave function to
have a node at $x=0$.
If now we go to the corresponding quantum trajectories, we observe
(see Fig.~\ref{fig9}(b), with red-solid line) that as the wave packet
starts to ``feel'' the presence of the wall, the trajectories bend
gradually (in the $x$ {\it vs} $t$ representation) and then start to
move in the opposite direction.
When these trajectories are compared with those associated with the
problem of the two wave-packet superposition, the resemblance between
trajectories with the same initial positions is excellent, except in
the interference region due to the different location of the nodes of
the corresponding wave functions ---these differences are the
trajectory counterpart of the shift mentioned before.

\begin{figure}
 \begin{center}
 \epsfxsize=14cm {\epsfbox{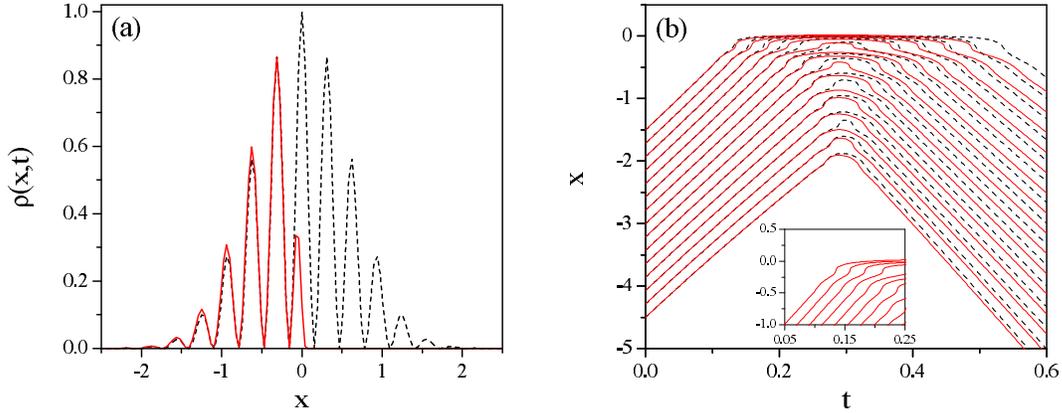}}
 \caption{\label{fig10}
  (a) Probability density at $t=0.3$ for the collision of a
  Gaussian wave packet off: the external potential described by
  Eq.~(\ref{extpot}) (red-solid line) and another (identical)
  Gaussian wave packet (black-dashed line).
  To compare, the maxima at $x \approx - 0.32$ of both probability
  densities are normalized to the same height (the maximum at $x=0$
  of the two wave packet probability density being set to unity).
  (b) Bohmian trajectories illustrating the dynamics associated with
  the two cases displayed in part (a).
  Inset: enlargement of the plot around $t=0.15$ to show the action of
  the potential well on the trajectories started closer to $x = 0$.
  Here, $v_p = 10$ and $v_s = 1$.}
 \end{center}
\end{figure}

From the previous description one might infer that, since the dynamics
for $x<0$ and for $x>0$ do not mix (due to non-crossing), each half of
the central interference maximum arises from different groups of
trajectories.
Thus, in principle, one should be able to arrange the impenetrable wall
problem in such a way that allows us to explain this effect.
Within this context, although all the peaks have the same width as in
the wave-packet superposition problem, the closest one to the wall
should have half such a width, i.e., $w \sim \pi\hbar/2p = w_0/2$.
Due to boundary conditions and the forward-backward interference
discussed above, it is clear that this peak cannot arise from
interference, but from another mechanism: a {\it resonance} process.
Therefore, apart from the wall, we also need to consider the presence
of a potential well.
In order to observe a resonance or quasi-bound state, the width
of this well should be, at least, of the order of the width $w$ of the
bound state.
From standard quantum mechanics, we know that in problems related to
bound states in finite well potentials the relationship \cite{schiff}
\be
 V_0 a^2 = n \frac{\hbar^2}{2m}
 \label{rel}
\ee
always appears, where $a$ is the half-width of the well ($a = w/2$) and
$n$ is an integer number.
The eventual solutions (bound states) are then observable or not
depending on whether the condition which it might correspond will
be in consonance or not with this condition.
In our case, we can use (\ref{rel}) to obtain an estimate of the well
depth, which results
\be
 V_0 = \frac{16}{\pi^2} \frac{p^2}{2m}
 \label{rel1}
\ee
when we assume $n=1$.
Now, we have then a potential which presents a short-range attractive
well before reaching the impenetrable wall,
\be
 V(x) = \left\{ \begin{array}{ccc}
   0      & \qquad & x < -w \\
  -V_0    & \qquad & -w \le x \le 0 \\
  \infty  & \qquad & 0 < x
  \end{array} \right. .
 \label{extpot}
\ee
If we compute now $\rho({\bf r},t)$ at $t^{\rm int}_{\rm max}$, we
obtain the result displayed in Fig.~\ref{fig10}(a).
As can be noticed, now there is an excellent matching of the peak
widths, with the closest one to the wall being half-width when compared
with the remaining ones ---the associate quantum trajectories are
displayed and compared in Fig.~\ref{fig10}(b).

\begin{figure}
 \begin{center}
 \epsfxsize=14cm {\epsfbox{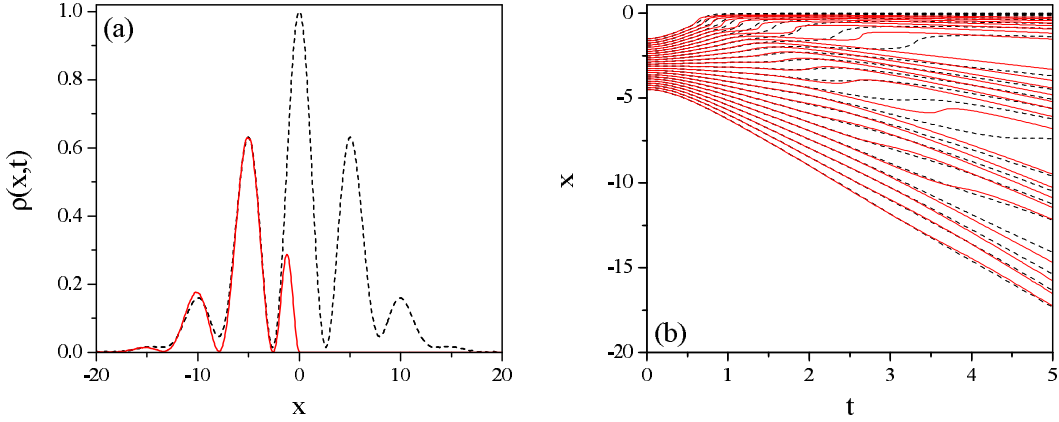}}
 \caption{\label{fig11}
  (a) Probability density at $t=5$ for the collision of a
  Gaussian wave packet off: the external, time-dependent potential
  described by Eq.~(\ref{caso8}) (red-solid line) and another
  (identical) Gaussian wave packet (black-dashed line).
  To compare, the maxima at $x \approx - 5$ of both probability
  densities are normalized to the same height (the maximum at $x=0$
  of the two wave packet probability density being set to unity).
  (b) Bohmian trajectories illustrating the dynamics associated with
  the two cases displayed in part (a).
  Here, $v_p = 0.1$ and $v_s = 1$.}
 \end{center}
\end{figure}

The equivalence between the two wave packet collisions and the
scattering of a wave packet off a potential is not restricted to the
condition $v_p > v_s$, but it is of general validity.
As shown in Fig.~\ref{fig6}, it also holds for the diffraction-like
situation, i.e., $v_p < v_s$.
In Fig.~\ref{fig11}(a), we show that half of the diffraction-like
pattern is again well reproduced after replacing one of the wave
packets by an external potential, and the same also happens for
the corresponding quantum trajectories (see Fig.~\ref{fig11}(b)).
However, in order to find these results, now a subtlety has to be
considered: the central diffraction maximum increases its width with
time.
In terms of simulating this effect with a potential function, it is
clear that the width of the potential well should also increase with
time.
Thus, we need to consider a ``dynamical'' or time-dependent potential
function rather than a static one, as done before.
In order to determine this potential function, we proceed as before.
First, we note that the two wave-packet collision problem,
Eq.~(\ref{sup1}), is explicitly written in terms of Gaussian wave
packets (see Eq.~(\ref{eqn2})) as
\be
 \Psi \sim
    e^{-(x+x_t)^2/4\tilde{\sigma}_t\sigma_0
     + ip(x+x_t)/\hbar + i Et/\hbar}
  + e^{-(x-x_t)^2/4\tilde{\sigma}_t\sigma_0
     - ip(x-x_t)/\hbar + i Et/\hbar} ,
 \label{caso1}
\ee
where
\be
 x_t = x_0 - v_p t
 \label{caso2}
\ee
(for simplicity, we have neglected the time-dependent prefactor,
since it is not going to play any important role regarding either
the probability density or the quantum trajectories).
The probability density associated with (\ref{caso1}) is
\be
 \rho(x,t) \sim
  e^{-(x+x_t^2)/2\sigma_t^2} + e^{-(x-x_t^2)/2\sigma_t^2}
  + 2 e^{-(x^2+x_t^2)/2\sigma_t^2} \cos \left[ f(t) x \right] ,
 \label{caso3}
\ee
with
\be
 f(t) \equiv \frac{\hbar t}{2m\sigma_0^2} \frac{x_t}{\sigma_t^2}
  + \frac{2p}{\hbar} .
 \label{caso4}
\ee
As can be noticed, (\ref{caso3}) is maximum when the cosine is $+1$
(constructive interference) and minimum when it is $-1$ (destructive
interference).
The first minimum (with respect to $x=0$) is then reached when
$f(t) x = \pi$, i.e.,
\be
 x_{\rm min}(t) = \frac{\pi}{\displaystyle \frac{2p}{\hbar}
  + \frac{\hbar t}{2m\sigma_0^2}\frac{x_t}{\sigma_t^2}}
  = \frac{\pi\sigma_t^2}{\displaystyle \frac{2p\sigma_0^2}{\hbar}
   + \frac{\hbar t}{2m\sigma_0^2} \ \! x_0} ,
 \label{caso5}
\ee
for which
\be
 \rho[x_{\rm min}(t)] \sim 4 e^{-(x_{\rm min}^2+x_t^2)/2\sigma_t^2}
   \sinh \left( \frac{x_{\rm min} x_t}{2\sigma_t^2} \right) ,
 \label{caso6}
\ee
which basically is zero if the initial distance between the two
wave packets is relatively large when compared with their spreading.

\begin{figure}
 \begin{center}
 \epsfxsize=14cm {\epsfbox{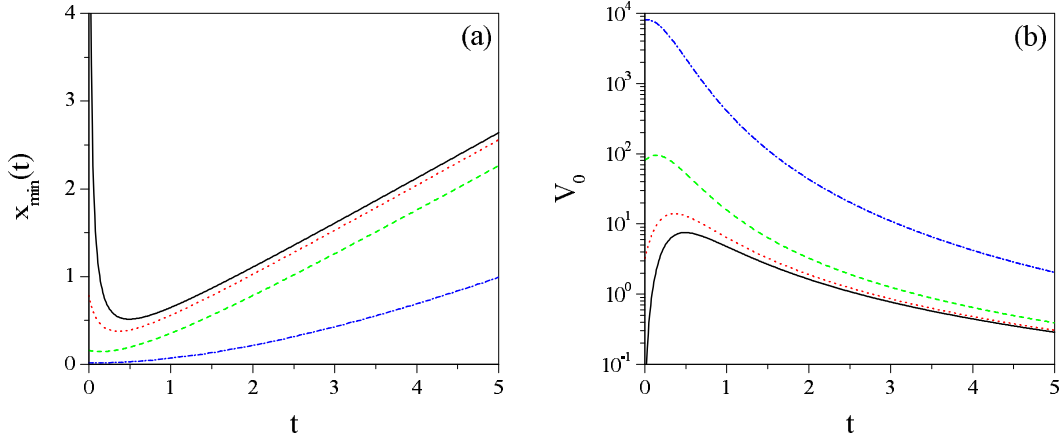}}
 \caption{\label{fig12}
  (a) Plot of $x_{\rm min}$ as a function of time for different values
  of the propagation velocity: $v_p = 0.1$ (black solid line),
  $v_p = 2$ (red dotted line), $v_p = 10$ (green dashed line) and
  $v_p = 100$ (blued dash-dotted line).
  (b) Plot of $V_0$ as a function of time for the same four values of
  $v_p$ considered in panel (a).
  In all cases, $v_s = 1$.}
 \end{center}
\end{figure}

In Fig.~\ref{fig12}(a), we can see the function $x_{\rm min}(t)$ for
different values of the propagation velocity $v_p$.
As seen, $x_{\rm min}(t)$ decreases with time up to a certain value,
and then increases again, reaching a linear asymptotic behavior.
From (\ref{caso5}) we find that the minimum value of $x_{\rm min}(t)$
is reached at
\be
 t_{\rm min} = \frac{4m\sigma_0^4}{\hbar^2 x_0}
  \left[ - p + \sqrt{p^2 + p_s^2 \left(\frac{x_0}{\sigma_0}\right)^2}
   \right] .
 \label{caso7}
\ee
The linear time-dependence at long times is characteristic of the
Fraunhofer regime, where the width of the interference peaks increases
linearly with time.
On the other hand, the fact that, at $t=0$, $x_{\rm min}(t)$ increases
as $v_p$ decreases (with respect to $v_s$) could be understood as a
``measure'' of the coherence between the two wave packets, i.e., how
important the interference among them is when they are far apart
(remember that, despite their initial distance, there is always an
oscillating term in between due to their coherence \cite{sanz-cpl2}).
Note that this is in accordance with the standard quantum-mechanical
arguments that interference-like patterns are manifestations of the
wavy nature of particles, while scattering-like ones display their
corpuscle nature (more classical-like).
Thus, as the particle becomes more ``quantum-mechanically'', the
initial reaching of the ``effective'' potential well should be larger.
And, as the particle behaves in a more classical fashion, this reaching
should decrease and be only relevant near the scattering or interaction
region, around $x=0$.
From Eq.~(\ref{caso5}), two limits are thus worth discussing.
In the limit $p \sim 0$,
\be
 x_{\rm min}(t) \approx \frac{\pi\sigma_t^2}{x_0} \frac{\tau}{t}
 \label{caso10}
\ee
and $t_{\rm min} \approx \tau$.
In the long-time limit, this expression becomes
$x_{\rm min}(t) \approx (\pi\hbar/2m)(t/x_0)$, i.e., $x_{\rm min}$
increases linearly with time, as mentioned above.
On the other hand, in the limit of large $\sigma_0$ (or, equivalently,
$v_p \gg v_s$),
\be
 x_{\rm min}(t) \approx \frac{\pi\hbar}{2p}
 \label{caso11}
\ee
and $t_{\rm min} \approx 0$.
That is, the width of the ``effective'' potential barrier remains
constant in time, this justifying our former hypothesis above, in the
scattering-like process, when we considered $w \sim \pi\hbar/2p$.

After Eq.~(\ref{caso5}), the time-dependent ``effective'' potential
barrier is defined as (\ref{extpot}),
\be
 V(t) = \left\{ \begin{array}{ccl}
   0      & \qquad & x < x_{\rm min}(t) \\
  -V_0[x_{\rm min}(t)]  & \qquad & x_{\rm min}(t) \le x \le 0 \\
  \infty  & \qquad & 0 < x
  \end{array} \right. .
 \label{caso8}
\ee
with the (time-dependent) well depth being
\be
 V_0[x_{\rm min}(t)] =
  \frac{2\hbar^2}{m} \frac{1}{x_{\rm min}^2(t)} .
 \label{caso9}
\ee
The variation of the well depth along time is plotted in
Fig.~\ref{fig12}(a) for the different values of $v_p$ considered
in Fig.~\ref{fig12}(b).
As seen, the well depth increases with $v_p$ (in the same way that
its width, $x_{\rm min}$, decreases with it) and decreases with time.
For low values of $v_p$, there is a maximum, which indicates the
formation of the quasi-bound state that will give rise to the innermost
interference peak (with half the width of the remaining peaks, as
shown in Fig.~\ref{fig11}(a)).
Note that, despite the time-dependence of the well depth, in the limit
$v_p \gg v_s$, we recover Eq.~(\ref{rel1}).

We have shown that the problem of the interference of two colliding
wave packets can be substituted by the problem of the collision of a
wave packet off a potential barrier.
Although the model that we have presented is very simple, it is
important to stress that it reproduces fairly well the dynamics
involved ---of course, one can always search for more refined and
precise models.
The fact that one can make this kind of substitutions puts quantum
mechanics and the superposition principle on the same grounds as two
important frameworks in classical physics (one could think of many
other situations, but these two ones are particularly general and
well known).
The first one is widely used in classical mechanics (and then, in its
quantized version, also in quantum mechanics): it is the equivalence
between a two-body problem and a one-body problem acted by a central
force.
As can be noticed, our reduction is of the same kind: a two wave-packet
dynamical process can be reduced to the dynamics of a single wave
packet subjected to the action of an external potential.
The second framework is the one provided by the so-called method of
images, widely used in electrostatics: the interaction between a
charge distribution and a conductor can be replaced by the interaction
between such a charge distribution and another virtual one.
This would be the reciprocal situation to ours: the dynamics induced
by an external potential on a wave packet can be translated as the
dynamics taking place when two wave packets (ours and a virtual one)
are considered.


\section{Final discussion and conclusions}
 \label{sec4}

Here we have presented a detailed analysis of quantum interference
within the framework of quantum trajectories.
As we have shown, there are at least two important properties which are
masked when interference is studied from the standard point of view,
but that become apparent as soon as we move to the Bohmian domain:
(a) it is always possible to determine uniquely the departure region
of a particle from the final outcome and (b) interference can be
described as an effect similar to that of particle-particle collisions.
Bohmian trajectories allow one to determine the origin of the particles
that contribute to each final wave packet, contrary to the standard
view of interference processes, where the final wave packets are
commonly associated with the time-evolved initial ones, according to
the superposition principle.
Nevertheless, as has also been shown here, putting aside the
superposition principle and considering the properties of the quantum
current density, it is also possible to reach similar conclusions in
standard quantum mechanics (although, to our knowledge, one cannot
find this in the literature).
Of course, although we might know that quantum particles are associated
with one wave packet or the other during the whole propagation, this
should not be mistakenly considered as to have a complete knowledge
of their corresponding initial positions in the experiment ---remember
that the initial positions distribute in a random fashion following the
initial probability density.

\begin{figure}
\begin{center}
\epsfxsize=13cm {\epsfbox{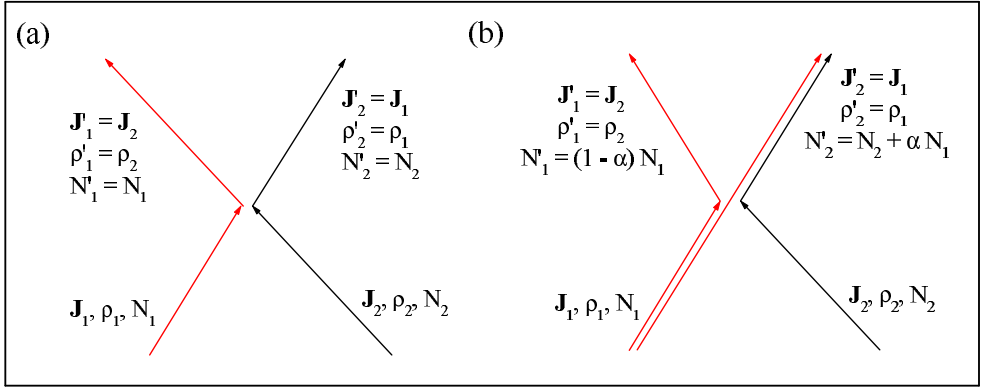}}
\caption{\label{fig13}
 Schematic representation of the two types of processes that may take
 place in two wave-packet collisions: (a) elastic-like scattering and
 (b) inelastic-like scattering. See text for details.}
\end{center}
\end{figure}

To some extent, the property (a) can then be regarded as a sort of
distinguishability, which should not be confused with that one
arising when one considers the problem of identical (fermion or
boson) particles.
Rather, this issue is connected with the property (b) in the sense that
it allows us to describe the process, in the collision-like case, as
classical collisions.
Two situations can then happen.
If the probability is equi-distributed between the two wave packets,
the effect is similar to classical elastic particle-particle
collisions, where only the momentum is transferred.
In our case, what is transferred is the probability distribution
through the probability density current, since the number of particles
associated with each wave packet remains the same.
This situation, represented schematically in Fig.~\ref{fig13}(a),
corresponds to the cases A and the first one of B described in
section~\ref{sec3.3}.
On the other hand, when the probabilities are asymmetrically
distributed, there is a probability transfer which translates into
a trajectory transfer ---the well defined non-crossing line then
disappears.
This behavior, which is similar to a classical inelastic
particle-particle collision, is displayed in Fig.~\ref{fig13}(b)
and corresponds to the case C and the second example of B (see
section~\ref{sec3.3}).
Rather than being just a conceptual discussion, it is worth noticing
that this analysis can be of much applied importance in the design,
development, improvement and/or implementation of new trajectory-based
algorithms \cite{wyatt-book}, where one of the main problems is
precisely the way to deal with separate (in space) wave packets and
the nodes emerging from their interference.

As has been noticed, observing collision-like behaviors or not depends
on the relation between the spreading and the propagation velocities of
the wave packets.
If $v_p$ is larger than $v_s$, one will observe collision-like
behaviors, which, to some extent, is the same to say that the wave
packets behave like classical particles or {\it corpuscles}.
Of course, note that when we use the concept ``collision'' we are not
referring exactly to a true particle-particle collision, since the two
wave packets indeed refer to the {\it same} system (which, by means of
any of the procedures mentioned in section~\ref{sec1}, for instance,
has been split up and converted into a so-called {\it Schr\"odinger
cat} system).
On the other hand, when $v_s$ dominates the dynamics, the behavior is
diffraction-like.
The wave packets then behave in a wavy manner and diffraction patterns
appear after their interference.

Finally, it is also very important the fact that wave-packet
interference problems can be understood within the context of
scattering off effective potential barriers.
As is well known, in classical mechanics one can substitute a
particle-particle scattering problem by that of an effective particle
(associated with the center of the mass of the particle-particle
system) acted by an (also effective) interaction central potential.
Similarly, here we have shown that any interference process
(either collision-like or diffraction-like) can also be rearranged in
such a way that the two wave-packet problem can be replaced by that of
a (single) wave-packet scattering off an effective (time-dependent)
potential barrier.
In this regard, we would like to note that this property could be
considered a precursor of more refined and well-known methods used to
deal with many body systems, where the many degrees of freedom are
replaced by a sort of effective time-dependent potential
\cite{sanz-dft}.
Furthermore, it is worth stressing that, in order to describe properly
the interference fringes during the process, we have shown that these
potentials have to support temporary bound states or resonances.
In those cases where the non-crossing boundary disappears because the
probability is not equally distributed between the two wave packets,
the impenetrable barriers should be replaced by ``transparent'' or step
barriers.
Thus, in the same way that after the collision process part of the
trajectories associated with one of the wave packets become attached to
the other wave packet, after the collision with the transparent barrier
part of the trajectories will move on the barrier.


\ack

This work has been supported by the Ministerio de Ciencia e
Innovaci\'on (Spain) under Project No.\ FIS2007-62006.
A.S.\ Sanz also acknowledges the Consejo Superior de Investigaciones
Cient\'{\i}ficas (Spain) for a JAE-Doc contract.


\end{document}